\documentclass[12pt]{article}
\usepackage{diagbox}
\usepackage{xcolor}
\usepackage{graphicx}
\usepackage{float}
\usepackage{amsthm,comment}
\usepackage{dsfont}
\usepackage{url}
\usepackage[ruled,vlined]{algorithm2e}
\usepackage{enumitem}
\RequirePackage{amsthm,amsmath,amsfonts,amssymb}
\RequirePackage[authoryear]{natbib}
\usepackage[ruled,vlined]{algorithm2e}
\usepackage{setspace}
\usepackage{tabularx} % in the preamble
\usepackage{booktabs}
\usepackage{authblk}
\usepackage{bbm}

\newtheorem{theorem}{Theorem}[section]
\newtheorem{lemma}[theorem]{Lemma}
\theoremstyle{remark}

\DeclareMathOperator*{\argmin}{argmin} % no space, limits underneath in displays
\RequirePackage[colorlinks,citecolor=blue,urlcolor=blue]{hyperref} 
\newcommand{\blind}{0}

\begin{document}

\def\spacingset#1{\renewcommand{\baselinestretch}%
{#1}\small\normalsize} \spacingset{0}

\spacingset{1}
\if1\blind
{
  \title{\bf Precision education: A Bayesian nonparametric approach for handling item and examinee heterogeneity in assessment data}
  \author[1]{Tianyu Pan}
  \author[1]{Weining Shen}
  \author[2]{Clintin P. Davis-Stober}
  \author[3]{Guanyu Hu}
  \affil[1]{Department of Statistics, University of California, Irvine}
  \affil[2]{Department of Psychological Sciences, University of Missouri - Columbia, Columbia, MO, 65211}
  \affil[3]{Department of Statistics, University of Missouri - Columbia, Columbia, MO, 65211}
  \date{}
  \maketitle
} \fi

\if0\blind
{
  \title{\bf Precision education: A Bayesian nonparametric approach for handling item and examinee heterogeneity in assessment data}
  \author[1]{Tianyu Pan}
  \author[1]{Weining Shen}
  \author[2]{Clintin P. Davis-Stober}
  \author[3]{Guanyu Hu}
  \affil[1]{Department of Statistics, University of California, Irvine}
  \affil[2]{Department of Psychological Sciences, University of Missouri - Columbia, Columbia, MO, 65211}
  \affil[3]{Department of Statistics, University of Missouri - Columbia, Columbia, MO, 65211}
  \date{}
  \maketitle
  \medskip
} \fi
\vspace{-.6cm}
%\bigskip
\spacingset{1.25}
\begin{abstract}
We propose a novel nonparametric Bayesian IRT model in this paper by introducing the clustering effect at question level and further assume heterogeneity at examinee level under each question cluster, characterized by the mixture of Binomial distributions. The main contribution of this work is threefold: (1) We demonstrate that the model is identifiable. (2) The clustering effect can be captured asymptotically and the parameters of interest that measure the proficiency of examinees in solving certain questions can be estimated at a $\sqrt{n}$ rate (up to a $\log$ term). (3) We present a tractable sampling algorithm to obtain valid posterior samples from our proposed model. We evaluate our model via a series of simulations as well as apply it to an English assessment 
data. This data analysis example nicely illustrates how our model can be used by test makers to distinguish different types of students and aid in the design of future tests.

\end{abstract}

\noindent%
\vfill

\noindent%
{\it } 
\vfill

\spacingset{1.5} % DON'T change the spacing!
\newpage
\section{Introduction}\label{sec: Intro}
Item response theory \citep[IRT;][]{mislevy1990modeling,rost1990rasch} was developed to better understand the mechanism behind an examinee passing a question (item) correctly and subsequently evaluate the discrepancies between examinees or items. The majority of IRT models assume that the response of an examinee to a question is probabilistic \citep{thomas2011value}, which is further determined by a latent parameter. For example, an accuracy parameter is within 0 and 1 if the response is True or False. In general, the latent probabilistic parameter relies on \textbf{two main factors}, i.e., the ability of an examinee and the difficulty of a question. Among all IRT models, the Rasch model \citep{rasch1993probabilistic} is one of the most well-known and widely used models thanks to its elegant format and interpretability. When being applied to a dichotomous matrix, where each column represents a test question and each row is an examinee's response, the Rasch model can be encapsulated as a logistic regression that models the $(i,j)$-th entry of the matrix using the linear predictor that contrasts the ability of the $i$-th examinee and the difficulty of the $j$-th question. While our proposed framework builds upon the Rasch model due to its simplicity, future work could consider extensions to other IRT frameworks, e.g., 2PL \citep{muraki1992generalized} and 3PL \citep{zumbo1997empirical,rouse1999advances}. We refer readers to \citet{thomas2011value} for a more comprehensive review on the history of IRT models.

Despite the usefulness of the IRT models, there is evidence that two key assumptions, \textit{unidimensionality} and \textit{local independence} \citep[Chapter 14;][]{andrich2019course}, may be violated in common applications \citep{keith1987functional,bell1988conditional,kreiner2007validity,marais2008formalizing}, that is, the ability of an examinee cannot be entirely represented by a single parameter (\textit{unidimensionality}) and the responses remain dependent after imposing the IRT model structure (\textit{local independence}). From a statistical point of view, violations of the these assumptions are inter-twined. When the mechanism behind the true data generating process cannot be recovered by using single ability or difficulty parameter that corresponds to an examinee or a question, it undoubtedly introduces conditional dependency on the responses due to lack-of-fit. Intuitively, unidimensionality is a proper assumption to unveil the Guttman pattern \citep{andrich2019course}, that is, the examinees who can correctly answer difficult questions with a certain probability should be able to answer easy questions with a higher probability. Nevertheless, this probably neglects the fact that some examinees could be more expert in handling certain questions than the others. Throughout this paper, we use the term \emph{superiority} to describe the phenomenon of an examinee being able to correctly answer classes of questions based on content type that is not captured by a uni-dimensional ability or difficulty parameter. A popular approach for accommodating multidimensional IRT data is to impose mixing structures on examinees' ability and questions' difficulty simultaneously, which generates a mixture of the Rasch models as a result. This idea has been investigated using frequentist \citep{rost1990rasch,alexeev2011spurious} and Bayesian \citep{bolt2002item,miyazaki2009bayesian,jang2018impact,sen2019model,hu2020nonparametric} approaches. Indeed, these methods enjoy more robustness and interpretability by introducing heterogeneity on examinees' ability and questions' difficulty, yet limited by the structure of the Rasch model, say these two main factors have to be linearly contrasted in each Rasch model mixture. Technically, using only the first order information of the two main factors without any interactions being considered, while ignoring possible interactions, can lead to lack-of-fit when the two main factors interact truly with one another in complex ways. \citet{bartolucci2017nonparametric} solves this problem by integrating the two main factors into a united accuracy parameter for each question, whilst assuming a global mixing structure on the accuracy parameters and letting them share a common sorting order over the mixing components. This can be interpreted as those examinees who have higher accuracy when answering questions of that same type. In addition, such construction introduces heterogeneity and dependency between the two main factors by quantifying an examinee's proficiency to a question using a mixture of accuracy parameters. 

However, none of the aforementioned works address the identifiability problem of a Binomial mixture distribution, which is crucial since it is not difficult to discover that the density functions are equivalent between $\text{Bernoulli}(0.5)$ and $0.5\times\text{Bernoulli}(0.1) + 0.5\times\text{Bernoulli}(0.9)$. If the model is not identifiable, researchers cannot expect a fast mixing when performing the Gibbs sampling, and even question the necessity of introducing a mixing structure, let alone ensure the consistency on the mixing parameters and interpret the results.

Motivated by the works introduced above, we propose a multidimensional IRT model based on nonparametric Bayesian procedure, which is termed as Averaged Constrained Binomial Mixture (ACBM). The objective of this paper is to relax the independence assumption intrinsically implied by the Rasch model, to model the \textit{superiority} phenomenon and to address the identifiability issue. Before we present our model, we could first imagine a heterogeneous pattern at both the examinee level and the question level that corresponds to the \textit{superiority} phenomenon, namely, examinees form several groups in each hypothetical cluster of the questions according to their proficiency in handling this type of questions, while the grouping pattern could differ across the question clusters. To model this idea, we consider the following steps. First, given a dichotomous response matrix, we aim to discover a partition over all questions such that in each question cluster, an examinees' responses to these questions can be characterized by a binomial distribution, of which the accuracy parameter, i.e., the examinee's proficiency in answering these questions, follows a mixing distribution. Technically, our proposed Bayesian model is essentially a prior over all mixing distributions given a partition on these questions, joint with a prior on all possible partitions on these questions. The key novelties of our proposed model are in interpretation and the theoretical guarantees of identifiability and posterior consistency. Our model can infer a partition on questions that reveals information at both levels. At the question level, in each question cluster, the examinees are automatically distinguished by their proficiency in tackling these questions using a mixing distribution. This allows us to discover a more complex accuracy patterns than just simple Guttman patterns \citep{andrich2019course}. At the examinee level, each mixing accuracy parameter under a given question cluster represents the proficiency of a specific group of examinees in handling these questions, which provides information for precision education. This, for example, provides statistical evidence to the test maker to identify the examinees who are not skilled in solving a certain question type. This information, in turn, is useful for designing and implementing additional questions within this type. In addition, the identifiability of our model is ensured by putting a dynamic upper bound on the number of mixing components in each question cluster, where the upper bound is determined by the size of the question cluster. We later address the identifiability of our model in Lemma \ref{1st-identifiability}. Our method can be tractably applied using an MCMC sampling algorithm for realization and is able to capture the true question partition and estimate the mixing parameters at a $\sqrt{n}$ rate (up to a $\log$ term), thanks to the rapid developments of Bayesian analysis over the past 30 years, in both computational \citep{neal2000markov,ishwaran2001gibbs} and theoretical \citep{nobile1994bayesian,shen2013adaptive,ho2016strong,ghosal2017fundamentals,guha2019posterior} research.

The rest of this article is organized as follows. The motivation and interpretation to our model is given in Section \ref{sec: Motivation}. The convergence results are presented in Section \ref{sec: Conv}, with the proof being deferred to the Supplementary File. In Section \ref{sec: Bayesian_inference}, we outline the MCMC sampling algorithm and introduce statistics to summarize the obtained posterior samples. The simulation study that validates our theoretical results and compares the performance between our model and the Rasch model is provided in Section \ref{sec: Simulation}. We carry out a real data analysis by applying our model to a test data set in Section \ref{sec: real data}. In Section \ref{sec: Discussion}, we discuss several possible ways to generalize our model, which paves the way for the future study. For ease of exposition, proofs, computation algorithms and
additional technical results are given in the supplementary materials.

\section{Motivation}\label{sec: Motivation}
Consider an $n$ by $D$ dichotomous matrix $\mathbf{X}$, whose $(i,j)$-th entry is a random variable, denoted by $X_{i,j}$, which takes value 1 if the $i$-th examinee answered the $j$-th question correctly and 0 otherwise, we let $\mathcal{D}\equiv\{1,2,\ldots,D\}$ be an index set on the total number of itmes, $\mathcal{C}$ be a partition of $\mathcal{D}$, and $c\in\mathcal{C}$ be a cluster of items defined by the partition $\mathcal{C}$. To illustrate the connection between $\mathcal{C}$ and $\mathcal{D}$ using an example, we take $D$ to be 3 and a partition $\mathcal{C}$ of $\mathcal{D}$ can be $\mathcal{C}=\{\{1,2\},\{3\}\}$. For this simple example, there are just two clusters, $c_{1}=\{1,2\}$ and $c_{2}=\{3\}$. Our model construction is motivated by the following example. Suppose a professor creates a test to evaluate her students, with the questions being classified into multiple clusters, with each cluster being an element $c \in \mathcal{C}$. Each question cluster $c$ can ideally distinguish different types of the students, of which the total number of types is given by $K^{(c)}$. Moreover, the $k$-th type indicates the proficiency of a specific group of students in answering a certain question type $c$, that is, the \textit{superiority}, which can further be quantified by an accuracy parameter $\theta^{(c)}_k$, for $1\leq k\leq K^{(c)}$. Inspired by this idea, we propose the following hierarchical order to model this structural clustering pattern,
\begin{equation}\label{ConsBinModel}
    \begin{split}
         & X_{1,j},\ldots,X_{n,j}\stackrel{i.i.d}{\sim} p^{(c)}_{F},~\text{for}~\forall j\in c,~\forall c\in\mathcal{C},\\
         & p^{(c)}_{F}(x)= \sum_{k=1}^{K^{(c)}}w^{(c)}_{k}\times (1-\theta^{(c)}_k)^{1-x}\times (\theta^{(c)}_k)^{x},~\text{for}~x=0~\text{or}~1,\\
         & (w^{(c)}_{1},\ldots,w^{(c)}_{K^{(c)}})\sim \text{Dir}(K^{(c)},(\alpha,\ldots,\alpha)),\\
         & \theta^{(c)}_k\stackrel{i.i.d}{\sim}\text{Beta}(a_0,b_0),~\text{for}~k=1,\ldots,K^{(c)},\\
         & K^{(c)}\sim q_0^{(c)}(K^{(c)};\gamma),\\
         & \mathcal{C}\sim m(\mathcal{C}),
    \end{split}
\end{equation}
where $\text{Dir}(K,(\alpha,\ldots,\alpha))$ refers to a Dirichlet distribution with $K$ categories and a concentration parameter $\alpha > 0$, $\text{Beta}(a_0,b_0)$ denotes a Beta distribution whose shape parameters are $a_0$ and $b_0$, $m(\mathcal{C})$ denotes a probability mass function over all possible partitions on $\mathcal{D}$, $q^{(c)}_0(\cdot;\gamma)$ denotes the Poisson distribution parameterized by $\gamma$ and truncated between $1$ and $(|c|+1)/2$ and $|\cdot|$, when being applied to a set $c$, refers to its cardinality instead of the absolute value when being applied to a real number. Note that the truncation on $q^{(c)}_0(\cdot;\gamma)$ refers to the maximum limit of a question cluster $c$ in dividing the students into different levels by their proficiency in answering a certain type of questions. This limitation relies on the question cluster size $|c|$. The term $(|c|+1)/2$ is the upper limit on the value $K^{(c)}$ and is a function of the number of items within the cluster $c$.  This upper limit can be interpreted as the ``resolution'' of the question cluster. The superscript $^{(c)}$ highlights that the mixing weights and the number of mixtures are allowed to vary across question clusters.

%Model \eqref{ConsBinModel} can be directly interpreted as, given a partition on the questions, we assume that within each question cluster $c$, an examinee's proficiency in answering these questions is entirely modeled by a mixture of Binomial distributions, whose $k$-th mixing weight and component are $w_{k}$ and $\text{Binomial}(|c|,\theta_{k}^{(c)})$ respectively. To illustrate this idea, we can consider an example that a professor creates a test to evaluate her students, with the questions being classified into several clusters. Each question cluster can ideally distinguish different types of the students, with each type indicating the proficiency of a specific group of students (group $k$ given $c$) in answering a certain type of questions (question cluster $c$), that is, the \textit{superiority}, which is quantified by the accuracy parameter $\theta_k^{(c)}$.

The main feature of our model lies in the dependency between the clustering structures at two levels. Specifically, the mixing distribution at the examinee-level is allowed to vary across question clusters, whereas most of the existing methods \citep{bolt2002item,miyazaki2009bayesian,jang2018impact,sen2019model,hu2020nonparametric} suggest that the mixing distribution at the examinee-level is invariant regardless of the heterogeneity at question-level. We conclude three main benefits for this feature of our model. Our model provides more interpretable results, say the heterogeneity in the examinees for each question type, compared with heuristically assuming mixing distributions on the two main factors separately, which can be hardly interpreted in this way. Such structural heterogeneity allows us to discover more complicated patterns than the Guttman pattern. For example, in a math test, it is reasonable to believe that some examinees are more proficient in algebra questions than they are in geometric questions. While some other examinees are more expert in solving geometric questions but not good at algebra questions. This cannot be explained by the Guttman mechanism, but can be justified by the \textit{superiority} phenomenon. In addition, we demonstrate that our model is identifiable when our chosen upper bound on the number of mixtures is imposed for each question cluster. Indeed, such an upper bound could inevitably lead to information loss when the size of a question cluster is small, but it intuitively makes sense because one cannot distinguish the proficiency of examinees to a certain type of questions only using very few of them. Moreover, the identifiability of our model further contributes to the identification of the true clustering structure on the questions, and meanwhile, provides $\sqrt{n}$ (up to a log term) estimates to the mixing weights and parameters in each question cluster under mild conditions.

In the next section, we proceed to present the technical details of our proposed method and introduce the posterior consistency results.

\section{Convergence results}\label{sec: Conv}

\subsection{Notations}\label{sec: notations}
In addition to the notations introduced in Section \ref{sec: Motivation}, throughout the rest of this paper, we will use $f(\cdot)$ or $f$ to denote a function $f(x)$ if it only takes a single argument. We let $X_i^{(c)}$ denote the $c$-section of random vector $X_i$, where $X_i$ refers to the dichotomous response of the $i$-th examinee. Similar definition is endowed to $x^{(c)}$ with respect to vector $\mathbf{x}$. We also let 
\begin{equation}\label{trueDensityUni}
    \begin{split}
         & p_{F_0}^{(c)}(x)=\sum_{k=1}^{K^{(c)}_0}w_{k,0}^{(c)}\times(1-\theta_{k,0}^{(c)})^{1-x}\times(\theta_{k,0}^{(c)})^x
    \end{split}
\end{equation}
be the true density function associated with a single question in cluster c. The corresponding true number of mixtures, mixing weights and component-wise parameters of $p_{F_0}^{(c)}(x)$ are hence $K_0^{(c)}$, $\{w_{k,0}^{(c)}\}_{k=1}^{K_0^{(c)}}$ and $\{\theta_{k,0}^{(c)}\}_{k=1}^{K_0^{(c)}}$. Analogously, a density function sampled following \eqref{ConsBinModel} is denoted by $p_F^{(c)}(x)$, whose number of mixtures, mixing weights and component-wise parameters are $K^{(c)}$, $\{w_{k}^{(c)}\}_{k=1}^{K^{(c)}}$ and $\{\theta_{k}^{(c)}\}_{k=1}^{K^{(c)}}$. We denote the true partition by $\mathcal{C}_0$ and any partition sampled following $m(\cdot)$ by $\mathcal{C}$. For random vector $X_i$, we let $p_{F_0}(\mathbf{x})$ and $p_F(\mathbf{x})$ be the true density function and a sampled density function following \eqref{ConsBinModel}, respectively. We also let $P_0$ and $\mathbb{P}_{F_0}$ be the probability distribution and the probability measure induced from $p_{F_0}$, and the expectation taken under $\mathbb{P}_{F_0}$ is denoted by $\mathbb{P}_{F_0}[f]$ or $\mathbb{P}_{F_0}f$.

For succinctness, we denote the prior on $p_F$ in \eqref{ConsBinModel} by $\Pi$ and let $\Pi^{(c)}$ for $\forall c\in\mathcal{C}$ be the prior on $p_F^{(c)}$ given $\mathcal{C}$. We define the set of all possible binomial mixtures as,

\begin{equation}\label{BinMixtures}
    \begin{split}
         & \mathcal{P}^{(c)}=\bigcup_{K=1}^{+\infty}\mathcal{P}^{(c)}(K)\equiv \left\{p^{(c)}_F(x^{(c)}):p^{(c)}_F(x^{(c)})=\sum_{k=1}^Kw^{(c)}_k(1-\theta_k^{(c)})^{|c|-\|x^{(c)}\|_1}\times (\theta_k^{(c)})^{\|x^{(c)}\|_1},\right.\\
         & \left. ~w^{(c)}_k\in(0,1),~\text{for}~k=1,\ldots,K,~\{\theta^{(c)}_{k}\}_{k=1}^K~\text{are distinct}\right\},
    \end{split}
\end{equation}
where the operator $\|\cdot\|_1$ refers to the $\ell_1$ norm when being applied to a vector, and in our situation, it is equivalent to a summation function as $x^{(c)}$ is a non-negative dichotomous vector. We point out that the prior $\Pi^{(c)}$ is essentially supported on $\mathcal{P}^{(c)}$ by truncating $K$ between 1 and $(|c|+1)/2$. For every $\mathcal{C}$ in the support of $m(\mathcal{C})$, we let $\Pi_{\mathcal{C}}\equiv \prod_{c\in\mathcal{C}}\Pi^{(c)}$ denote the prior on $\prod_{c\in\mathcal{C}}p_F^{(c)}$. It follows that

\begin{equation}\label{avgPrior}
    \begin{split}
         & \Pi = \sum_{\mathcal{C}\in\mathcal{S}}m(\mathcal{C})\Pi_{\mathcal{C}},
    \end{split}
\end{equation}
where $\mathcal{S}$ refers to the collection of all possible partitions of $\mathcal{D}$. For every $c\in\mathcal{C}$ and every $\mathcal{C}\in\mathcal{S}$, we use $p_{F_0}^{(c)}(x^{(c)})\equiv \int p_{F_0}(\mathbf{x})d x^{(\mathcal{D}\setminus c)}$ to denote the marginal density of $p_{F_0}$ on the sub-vector $x^{(c)}$ and $p_{F}^{(c)}(x^{(c)})$ to denote the joint density function on $x^{(c)}$ given $p_F^{(c)}(x)$, with similar definition as \eqref{trueDensityUni}.

\subsection{Convergence results}\label{sec: consistency}
We begin with the interpretations to the following three assumptions,

\begin{enumerate}
    \item[(A1)] The true partition $\mathcal{C}_0$ is in the support of $m(\mathcal{C})$.
    \item[(A2)] For $\forall c\in\mathcal{C}_0$, it holds for $p_{F_0}^{(c)}$ that 
    \begin{equation}\label{A2}
    \begin{split}
         & \emph{\text{Distinctness:}}~0 < \theta_{i,0}^{(c)}\neq\theta_{j,0}^{(c)} < 1,~\text{for}~1\leq i\neq j\leq K_0^{(c)},\\
         & \emph{\text{Non-trivial weights:}}~w_{k,0}^{(c)}>0,~\text{for every}~1\leq k\leq K_0^{(c)},\\
         & \emph{\text{Bounded component number:}}~K_0^{(c)}\leq \frac{|c|+1}{2}.
    \end{split}
    \end{equation}
    \item[(A3)] For every $\mathcal{C}$ in the support of $m(\mathcal{C})$, the true density has at least $\epsilon_0 > 0$ distance away from the best estimation induced by $\mathcal{C}$ with respect to the Kullback–Leibler divergence\footnote{The Kullback–Leibler divergence between $f(x)$ and $g(x)$ is defined as 
$$\text{KL}(f(\mathbf{x});g(\mathbf{x})) = \int\log\left(\frac{f(\mathbf{x})}{g(\mathbf{x})}\right)f(\mathbf{x})d\mathbf{x}.$$}, that is,
    $$\left\|\prod_{c\in\mathcal{C}_0}p_{F_0}^{(c)}(x^{(c)})-\prod_{c\in\mathcal{C}}p_{F^*}^{(c)}(x^{(c)})\right\|_1 > \epsilon_0,$$ where $p_{F^*}^{(c)}(x^{(c)})$, for $\forall c\in\mathcal{C}$ is obtained by minimizing $\text{KL}\left(p_{F_0}^{(c)}(x^{(c)});p_{F}^{(c)}(x^{(c)})\right),$ with respect to $p_{F}^{(c)}(x^{(c)})\in\mathcal{P}^{(c)}$.
\end{enumerate}
The first two assumptions are quite standard. Assumption (A1) is necessary to ensure that our model is correctly specified. The only eye-catching part in Assumption (A2) is the constraint on the number of components under each question cluster. This constraint aims to guarantee that the binomial mixture part in our proposed model is identifiable, as noted by \citet{teicher1963identifiability}. Later in Lemma \ref{1st-identifiability}, we will prove that the binomial mixture model is identifiable at the first-order. Assumption (A3) is a weaker condition than the general identifiability assumption. In fact, the density function $\prod_{c\in\mathcal{C}}p^{(c)}_F(x^{(c)})$ is non-identifiable for some instances of $\left\{\{w_k^{(c)},\theta_k^{(c)}\}_{c\in\mathcal{C}},\mathcal{C}\right\}$, for example, the corresponding density functions are identical given different parameterizations $$\{\{w_1^{c} = 1, \theta_1^{(c)} = 0.5\}_{c = \{1,2\}},\{w_1^{c} = 1, \theta_1^{(c)} = 0.5\}_{c = \{3\}}\}$$ and $\{\{\theta_1^{(c)} = 1,\theta_1^{(c)} = 0.5\}_{c = \{1,2,3\}}\}$ when $D = 3$. Therefore, one cannot directly apply Doob's theorem \citep{doob1949application} but need to resort to Assumption  (A3) and Theorem \ref{selectionConsistency} for the consistency on $\mathcal{C}$.

\begin{theorem}\label{selectionConsistency}
Assume (A1) and (A3) are satisfied, then it holds that
\begin{equation}\label{eqSelectionConsistency}
    \begin{split}
     & \Pi(\mathcal{C}=\mathcal{C}_0\mid X_{1},\ldots,X_{n})\to 0~~a.s.~P_0.
    \end{split}
\end{equation}

\end{theorem}

Theorem \ref{selectionConsistency} states that our proposed model can correctly identify the latent question partition asymptotically, that is, when sample size increases, we can expect that the question clustering configuration sampled from the posterior distribution of our model will eventually converge to the true question partition. 

To pave the way to our next theorem, we prove that the binomial mixture model is identifiable at the first order, which is defined as follows. 

\begin{lemma}\label{1st-identifiability}
(First-order Identifiability) Assume that $\{\theta_i\}_{i=1}^K$ are distinct and $\{\alpha_i\}_{i=1}^K$ and $\{\beta_i\}_{i=1}^K$ are real-valued coefficients and (A3) holds. Suppose that
\begin{equation}\label{1stIdent}
    \begin{split}
        & \sum_{i=1}^K \alpha_if(y\mid,\theta_i,n)+\sum_{i=1}^K\beta_i \frac{\partial f}{\partial \theta_i}f(y\mid,\theta_i,n)= 0,
    \end{split}
\end{equation}
where $f(y\mid,\theta_i,n) = {n\choose y}(1-\theta_i)^{n-y}(\theta_i)^y$ and 
$$\frac{\partial f}{\partial \theta_i}f(y\mid,\theta_i,n) = \mathbbm{1}(1\leq y\leq n)n{n-1\choose y-1}(1-\theta_i)^{n-1-(y-1)}(\theta_i)^{y-1} - $$
$$\mathbbm{1}(0\leq y\leq n-1)n {n-1\choose y}(1-\theta_i)^{n-1-y}(\theta_i)^{y},$$ then it implies $\alpha_1=\beta_1=\ldots = \alpha_K=\beta_K=0$, if $1\leq K \leq (n+1)/2$.

\end{lemma}

Lemma \ref{1st-identifiability} is indispensable for estimating, at a $\sqrt{n}$ rate, the true mixing weights and the true component-wise parameters under each question cluster. The results of Theorem \ref{selectionConsistency} and Lemma \ref{1st-identifiability} yield to our final result.

\begin{theorem}\label{postContract}
Assume (A1)-(A3) are satisfied, then the proposed model can estimate the true parameters a posteriori, given a contraction rate $\epsilon_n$,
\begin{equation}\label{eqPostContract}
    \begin{split}
     & \Pi(\{p_F^{(c)}: |w_{\sigma^{(c)}(i)}^{(c)}-w_{i,0}^{(c)}|\lesssim M'\epsilon_n,\|\theta_{\sigma^{(c)}(i)}^{(c)}-\theta_{i,0}^{(c)}\|_2\lesssim M'\epsilon_n,~\text{for}~i=1,\ldots,K^{(c)},\\ 
     & K^{(c)}=K_0^{(c)},\forall c\in\mathcal{C},\mathcal{C}=\mathcal{C}_0\}\mid X_{1},\ldots,X_{n})\to 1,~~a.s.~P_0,
    \end{split}
\end{equation}
where $M$ is a large positive constant to be determined and $M' > 0$ is a universal constant that only depends on $M$, $\epsilon_n = (\log(n))^t/\sqrt{n}$, for any $t > 1$.

\end{theorem}

Theorem \ref{postContract} indicates that our model can detect the heterogeneity in examinees, and meanwhile, identify the true question partition.

\section{Bayesian inference and algorithm}\label{sec: Bayesian_inference}

We begin with the outline of our posterior sampling algorithm, which consists of the following four steps,
\begin{itemize}
    \item (1) For each column (question), calculate the marginal likelihood when all students possess the same accuracy in answering this question, say marginalizing the Binomial likelihood over the Beta prior in \eqref{ConsBinModel}.
    \item (2) Conditioning on the row (examinee) assignment under each column (question) cluster, update the column assignment by enumerating from column 1 to column $D$.
    \item (3) Conditioning on the column (question) assignment, update the row (examinee) assignment under each column (question) cluster by enumerating from row 1 to row $n$ for $n_{rep}$ times.
    \item (4) Loop between (2) and (3) for $n_{iter}$ times to approach the stationary distribution.
\end{itemize}
The detailed algorithm is deferred to the Supplementary File. It is worthwhile to point out that Step (2) is essentially the Algorithm 1 proposed by \citet{neal2000markov}, by treating each column (question) to be an ``individual", whose parameter is the row-wise partition on the examinees. To update the row-wise partition in Step (3), we modify the aforementioned Algorithm 1 to accommodate the upper bound on the number of (student) mixtures under each column (question) cluster, which leads to a sampling scheme which is slightly different from Theorem 4.1 of \citet{miller2018mixture}. Based on the findings in our simulation studies, we notice that $n_{iter}=200$ and $n_{rep}=400$ is sufficient to obtain trustworthy posterior samples when $n\leq 1000$ and $D\leq 80$. In addition, for the hyper-parameters defined in \eqref{ConsBinModel}, we let both $a_0$ and $b_0$ be $.01$ and $\gamma$ equal to 1 such that the prior information is sufficiently non-informative but new clusters still have enough probability to be generated for both column (question) and row (examinee) in practice. The probability mass function $m(\mathcal{C})$ is chosen as the exchangeable partition probability function \citep[EPPF;][]{pitman2002combinatorial} of the mixture of finite mixtures model \citep[MFM;][]{miller2018mixture} to ensure a closed form on the full conditional distribution when sampling. The most time consuming task ($n=1,000$ and $D=80$) among the simulation studies and the real data analysis takes approximately 6 hours to finish after being assigned to a server with 94.24GB RAM, 24 processing cores, operating at 3.33GHz.

To summarize the posterior samples, we use the following three statistics to estimate (1) the column (question) partition, (2) the row-wise (examinee) partition under each column cluster and (3) the component-wise accuracy under each column cluster. The column (question) partition is estimated using Dahl's estimate \citep{dahl2006model}, defined as follows,

\begin{equation}\label{ColumnEst}
    \begin{split}
        & \hat{\ell}=\argmin_{1\leq \ell\leq M}\sum_{i=1}^n\sum_{j=1}^n\left\{\delta_{i,j}(\mathcal{C}^{\text{Col}}(\ell))-\hat{\pi}^{\text{Col}}_{i,j}\right\}^2,\\
        & \hat{\mathcal{C}}^{\text{Col}} = \mathcal{C}^{\text{Col}}(\hat{\ell}),
    \end{split}
\end{equation}
where $M$ is the number of MCMC iterations after burn-in, $\mathcal{C}^{\text{Col}}(\ell)$ refers to the column assignment at the $\ell$-th iteration after burn-in, $\delta_{i,j}(\mathcal{C}^{\text{Col}}(\ell))$ is an indicator function, defined as $\mathbbm{1}(\mathcal{C}^{\text{Col}}_i(\ell) = \mathcal{C}^{\text{Col}}_j(\ell))$, with $\mathcal{C}^{\text{Col}}_i(\ell)$ denoting the clustering assignment of the $i$-th column, and $\hat{\pi}^{\text{Col}}_{i,j}$ is obtained by averaging $\delta_{i,j}(\mathcal{C}^{\text{Col}}(\ell))$ over post burn-in MCMC samples, namely, $ \hat{\pi}^{\text{Col}}_{i,j} =\frac{1}{M}\sum_{\ell=1}^M\delta_{i,j}(\mathcal{C}^{\text{Col}}(\ell))$. The column partition summarized by Dahl's estimate is believed to be the most representative one as it minimizes the entry-wise $\ell_2$-distance between the self-concordance matrix of a given partition and the probability matrix $\hat{\pi}^{\text{Col}}_{i,j}$ that any pair of columns $i$ and $j$ being clustered together. The row-wise (student) partition is then summarized from the iterations where the column (question) partition is equal to $\mathcal{C}^{\text{Col}}(\hat{\ell})$,
\begin{equation}\label{RowEst}
    \begin{split}
        & \hat{\ell} =\argmin_{1\leq \ell\leq M; \mathcal{C}^{\text{Col}}(\ell) = \hat{\mathcal{C}}^{\text{Col}}}\sum_{d=1}^D\sum_{i=1}^n\sum_{j=1}^n\left\{\delta_{i,j}(\mathcal{C}^{\text{Row};d}(\ell))-\hat{\pi}^{\text{Row};d}_{i,j}\right\}^2,\\
        & \hat{\mathcal{C}}^{\text{Row};d} = \mathcal{C}^{\text{Row};d}(\hat{\ell}),\text{ for }d=1,\ldots, D,
    \end{split}
\end{equation}
where $\mathcal{C}^{\text{Row};d}(\ell)$ refers to the row assignment of the $d$-th column at the $\ell$-th iteration after burn-in and $\hat{\pi}^{\text{Row};d}_{i,j}$ is defined in a similar way with $\hat{\pi}^{\text{Col}}_{i,j}$ for the $d$-th column. It can be expected that ties happen for $\hat{\mathcal{C}}^{\text{Row};d}$ given $\forall d\in c$ and $\forall c\in\hat{\mathcal{C}}^{\text{Col}}$ by definition. Analogous to the idea behind $\hat{\mathcal{C}}^{\text{Col}}$, $\hat{\mathcal{C}}^{\text{Row};d}$ seeks for an iteration such that the squared $\ell_2$ distance is minimized averaged over all columns. The component-wise accuracy under each column cluster is then estimated using a posterior mean given $\hat{\mathcal{C}}^{\text{Col}}$ and $\hat{\mathcal{C}}^{\text{Row};d}$ for $d=1,\ldots,D$.

\section{Simulation}\label{sec: Simulation}

We study our proposed method using four data generating processes (DGP) and compare the result of our model with the one given by the Rasch model \citep{rasch1993probabilistic}. The Rasch model is realized using \textit{tam} package in R. The four DGPs are designed to mimic the situations when the data are generated under our proposed model or the Rasch model, given increasing number of examinees ($n=100,300,1,000$). We proceed by outlining the first four DGPs and defer the details to the Supplementary File. The first two DGPs are designed under our model,
\begin{itemize}
    \item \textbf{DGP1}. 20 questions are divided into 5 column (question) clusters, with 3 large question clusters and the rest 2 questions individually forming two question clusters. Under each column (question) cluster, the accuracy within each mixture stays identical and the mixture number satisfies the constraint.
    \item \textbf{DGP2}. 60 questions are divided into 5 column (question) clusters, with 3 large question clusters and the rest 2 questions individually forming two question clusters. Under each column (question) cluster, the accuracy within each mixture stays identical and the mixture number satisfies the constraint.
\end{itemize}

The last two DGPs are generated following the Rasch model, that is,
\begin{equation}\label{RaschModel}
    \begin{split}
    & X_{i,j}\stackrel{\text{ind}}{\sim} \text{Bernoulli}(\theta_{i,j}),~\text{for}~i=1,\ldots,n~\text{and}~j=1,\ldots,D,\\
    & \theta_{i,j} = \frac{\exp\{\xi_i-\psi_j\}}{1+\exp\{\xi_i-\psi_j\}},
\end{split}
\end{equation}
with the two DGPs being presented as follows,
\begin{itemize}
    \item \textbf{DGP3}. 20 questions are divided into 2 column (question) clusters by letting $\psi_j$ take value from $(-0.5,0.5)$ and 3 row clusters by letting $\xi_i$ randomly take value from $(-2,0,2)$, following the data generating process defined in \eqref{RaschModel}.
    \item \textbf{DGP4}. 60 questions are divided into 2 column (question) clusters by letting $\psi_j$ take value from $(-0.5,0.5)$ and 3 row clusters by letting $\xi_i$ randomly take value from $(-2,0,2)$, following the data generating process defined in \eqref{RaschModel}.
\end{itemize}

To validate Theorem \ref{postContract}, we consider the first two DGPs and adopt the following criteria 
\begin{equation}\label{trueCriteria}
    \begin{split}
        & \text{CWRI} = \text{RI}(\hat{\mathcal{C}}^{\text{Col}},\mathcal{C}_0),\\
        & \text{ADK} = \frac{1}{D}\sum_{d=1}^D\left||\hat{\mathcal{C}}^{\text{Row};d}|-K^{(c(d))}_0\right|,\\
        & \text{ADW} = \mathrm{1}(\text{RI}(\hat{\mathcal{C}}^{\text{Col}},\mathcal{C}_0) = 1)\times \frac{1}{|\mathcal{C}_0|}\sum_{c\in\mathcal{C}_0}\frac{1}{K_0^{(c)}}\min_{\sigma^{(c)}}\sum_{i=1}^{K^{(c)}_0}\left|\hat{w}^{(c)}_{\sigma^{(c)}(i)}-w^{(c)}_{0;i}\right|+\\
        & \mathrm{1}(\text{RI}(\hat{\mathcal{C}}^{\text{Col}},\mathcal{C}_0) \neq 1) \times 2,\\
        & \text{ADP} = \mathrm{1}(\hat{K}^{(c)} \geq K^{(c)}_0,c\in \mathcal{C}_0)\times \frac{1}{|\mathcal{C}_0|}\sum_{c\in\mathcal{C}_0}\sqrt{\frac{1}{K_0^{(c)}}\min_{\sigma^{(c)}}\sum_{i=1}^{K^{(c)}_0}\left\|\hat{\theta}^{(c)}_{\sigma^{(c)}(i)}-\theta^{(c)}_{0;i}\right\|_2^2}+\\
        & \left(\mathrm{1}(\hat{K}^{(c)} < K^{(c)}_0) \vee \mathrm{1}(c\notin \mathcal{C}_0))\right)\times 1,
    \end{split}
\end{equation}
where $\text{CWRI}$, $\text{ADK}$, $\text{ADW}$ and $\text{ADP}$ are the abbreviations of column-wise Rand Index, averaged absolute difference in the row-wise number of component, averaged absolute difference in the row-wise weights and averaged $\ell_2$ difference in the row-wise accuracy, $\text{RI}(\mathcal{C},\mathcal{C}')$ denotes the Rand Index \citep{rand1971objective} between $\mathcal{C}$ and $\mathcal{C}'$, $c(d)$ represents the column cluster $c$ where the $d$-th column is assigned to, $\sigma^{(c)}(\cdot)$ refers to the permutation operator, $w_{0;i}^{(c)}$ and $\theta_{0;i}^{(c)}$ denote the $i$-th true mixing weight and the $i$-th true component-wise accuracy under the column (question) cluster $c$ respectively. $\hat{w}_{i}^{(c)}$ and $\hat{\theta}_i^{(c)}$ represent the estimated values of $w_{0;i}^{(c)}$ and $\theta_{0;i}^{(c)}$ using posterior mean. Note that the penalty of misidentifying the true column (question) partition is added to $\text{ADW}$, which matches the maximum difference between the estimated mixing weights and the true mixing weights. Similar penalty is also attached to $\text{ADP}$. Ideally, we expect $\text{CWRI}$ converges to 1 and the rest three criteria shrink toward 0 if Theorem \ref{postContract} is true. The correct limiting values and the decreasing standard error successfully manifest our theoretical results, 
suggested by Table \ref{tab:1}.

\begin{table}[htp]
    \centering
        \caption{Median (Standard error) of the four criteria over 100 Monte Carlo replications for each of the first two DGPs given different sample sizes.}
    \label{tab:1}
   \begin{tabular}{l|l|l|l|l|l}
   \toprule
    \toprule
DGP  & n & CWRI & ADK & ADW & ADP\\
    \midrule
  & 100 & 1.000 (0.005) & 0.000 (0.181) & 0.082 (0.395) &  0.064 (0.192)\\
1 & 300 & 1.000 (0.000) & 0.000 (0.158) & 0.034 (0.015) & 0.033 (0.014) \\
  & 1000 & 1.000 (0.000) & 0.000 (0.097) & 0.014 (0.006) & 0.013 (0.007)\\
    \midrule
  & 100 & 1.000 (0.001) & 0.625 (0.273) & 0.811 (0.338) & 0.417 (0.166) \\
2 & 300 & 1.000 (0.000) & 0.000 (0.144) & 0.029 (0.136) & 0.023 (0.068) \\
  & 1000 & 1.000 (0.000) & 0.000 (0.075) & 0.016 (0.005) & 0.013 (0.003) \\
  \bottomrule
  \bottomrule
    \end{tabular}
\end{table}

In the last two DGPs, the assumptions of the Rasch model are satisfied. We propose to study the performance of our model in identifying the true column (question) and row (student) partitions, defined as the labelling of $\psi_j$ and $\xi_i$ respectively, and compare the performance of estimating $\theta_{0;i}^{(d)}\equiv \frac{\exp\{\xi_i-\psi_j\}}{1 + \exp\{\xi_i-\psi_j\}}$ using the following two criteria in addition to $\text{CWRI}$,

\begin{equation}\label{twoCriteria}
    \begin{split}
        & \text{ARWRI} = \frac{1}{D}\sum_{d=1}^D\text{RI}(\hat{\mathcal{C}}^{\text{Row};d},\mathcal{C}^{\text{Row;(c(d))}}_0),\\
        & \text{D}_1 = \frac{1}{nD}\sum_{i=1}^n\sum_{d=1}^D\left|\hat{\theta}_{i}^{(d)}-\theta_{0;i}^{(d)}\right|
    \end{split}
\end{equation}
where ARWRI is the abbreviation of averaged row-wise Rand Index and $\hat{\theta}_{i}^{(d)}$ can be directly provided by the Rasch model or using the posterior mean for our model. The results are presented in Table \ref{tab:2},

\begin{table}[htp]
    \centering
        \caption{Median (Standard error) of the three criteria over 100 Monte Carlo replications for each of the last two DGPs given different sample sizes.}
    \label{tab:2}
   \begin{tabular}{l|l|l|l|l|l}
   \toprule
    \toprule
\multicolumn{2}{c|}{} & \multicolumn{3}{c|}{ACBM} & Rasch\\
    \midrule
    \midrule
DGP  & n & CWRI & ARWRI & $\text{D}_1$ & $\text{D}_1$\\
    \midrule
  & 100 & 0.474 (0.193) & 0.922 (0.088) & 0.093 (0.018) & 0.073 (0.005)\\
3 & 300 & 1.000 (0.048) & 0.814 (0.020) & 0.077 (0.015) & 0.068 (0.003)\\
  & 1000 & 1.000 (0.000) & 0.818 (0.009) & 0.066 (0.005) & 0.066 (0.002)\\
\midrule
  & 100 & 0.919 (0.174) & 0.978 (0.016) & 0.026 (0.022) & 0.050 (0.003)\\
4 & 300 & 1.000 (0.007) & 0.979 (0.007) & 0.012 (0.003) & 0.043 (0.002)\\
  & 1000 & 1.000 (0.000) & 0.977 (0.004) & 0.009 (0.001) & 0.040 (0.001)\\
  \bottomrule
  \bottomrule
    \end{tabular}
\end{table}
It is interesting to note that when $n$ increases, $\text{CWRI}$ increases towards 1 and $\text{ARWRI}$ stays at a high value. Though $\text{ARWRI}$ is not guaranteed to converge towards 1, our model is able to identify most of the correct labels for examinees when the latent accuracy parameters are sufficiently well separated. In addition, our model achieves a higher $\text{ARWRI}$ when more questions are available under each question cluster, which matches our intuition, that is, more questions are more helpful in correctly distinguishing different types of students, by comparing the results of DGP4 with the ones of DGP3. By comparing the $\text{D}_1$ values of our proposed model and the Rasch model, our model provides a more efficient estimate to the accuracy parameter $\theta_{0,i}^{(c)}$, especially when $n$ and $D$ are large, say DGP4.

\section{Test Data Analysis}\label{sec: real data}
\subsection{Descriptive analysis}\label{sec: descriptive}
\begin{figure}[htp]
\minipage{0.48\textwidth}
  \includegraphics[width=\linewidth]{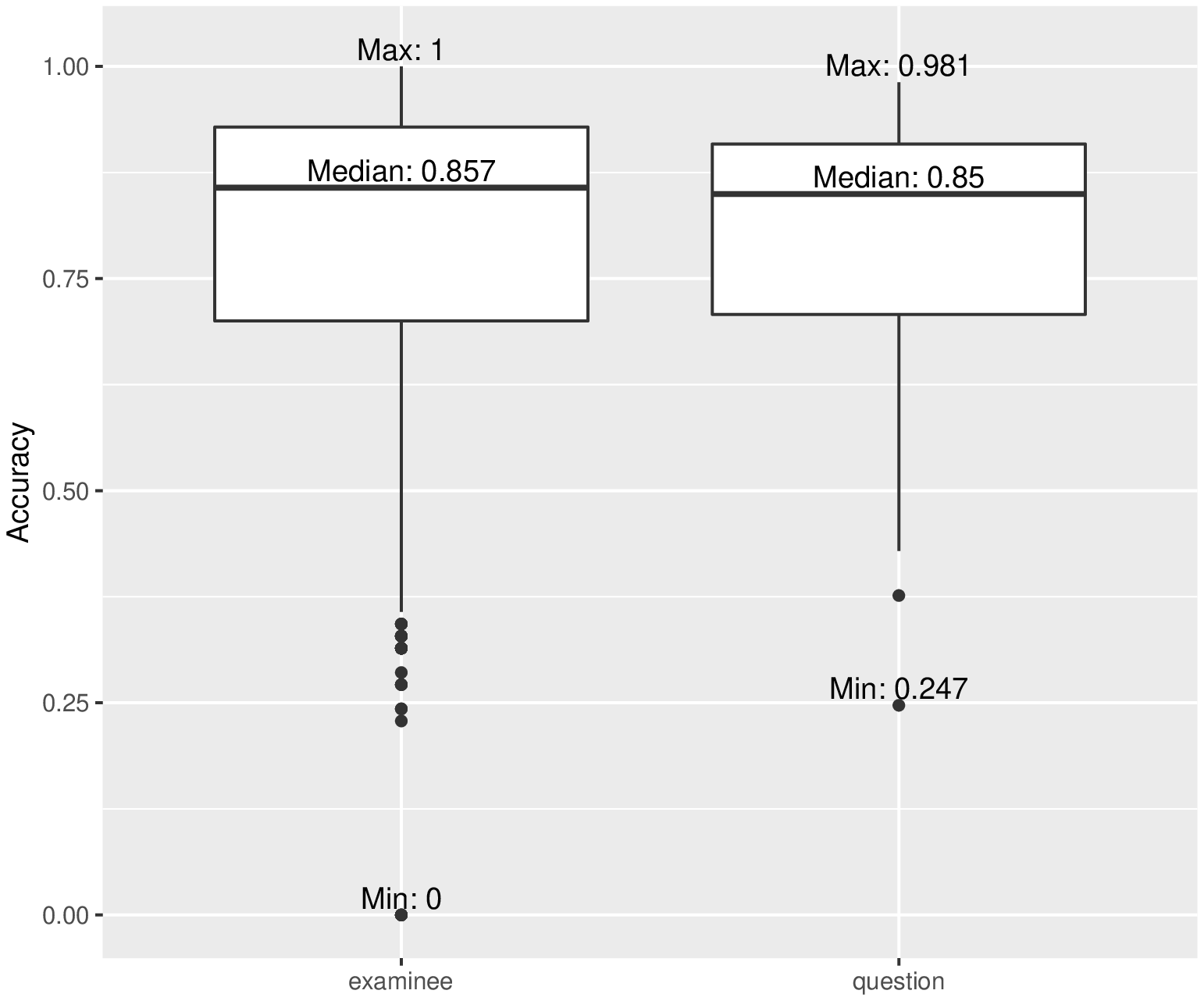}
\endminipage\hfill
\minipage{0.48\textwidth}
  \includegraphics[width=\linewidth]{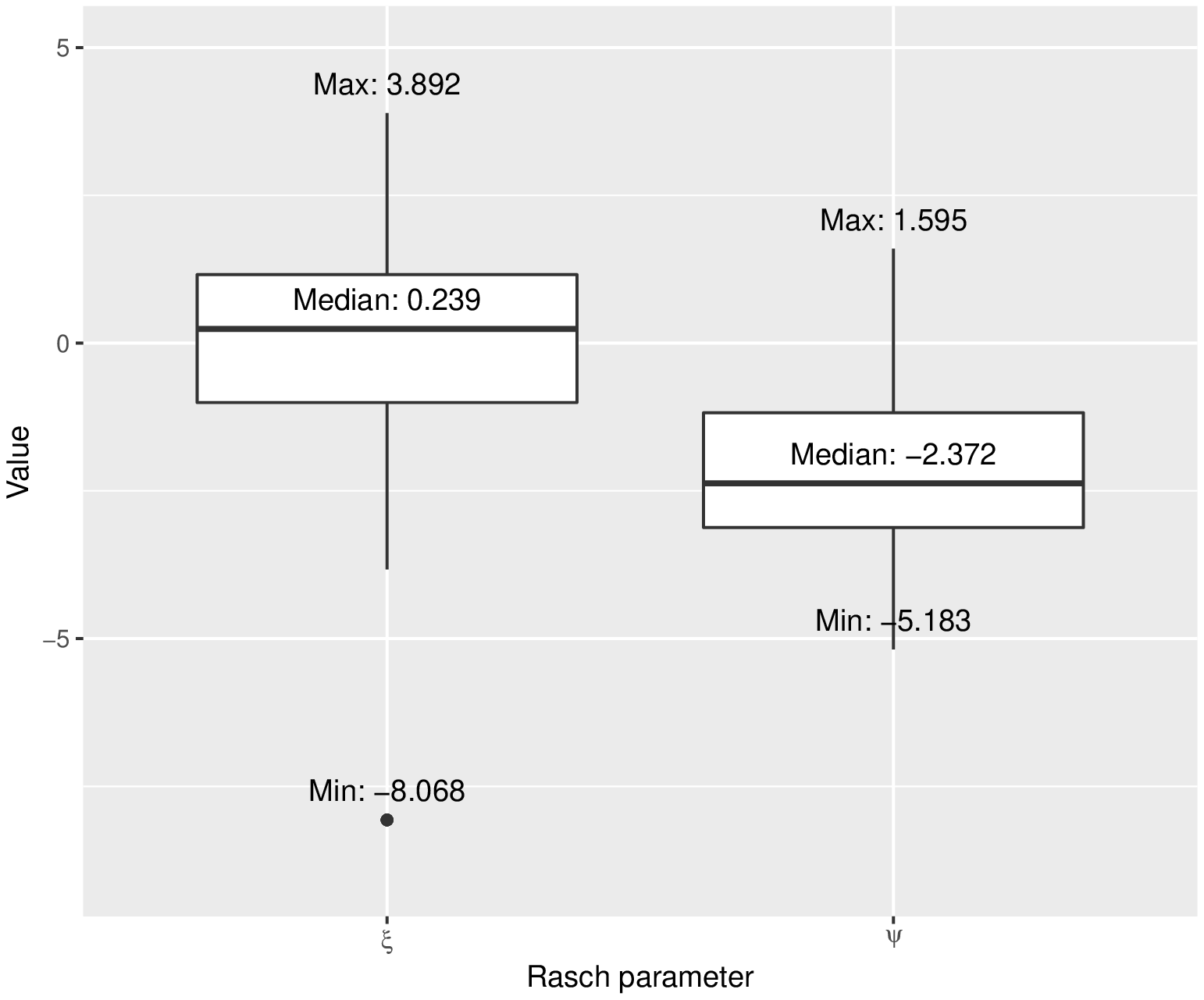}
\endminipage\hfill
\caption {\label{fig: expAnalysis} Left: the boxplots of the estimated accuracy marginalized over each question or examinee; Right: the boxplots of the Rasch parameters.}
\end{figure}

The motivating data consist of the English exam results of the 2020–2021 academic year from No. 11 Middle School of Wuhan, Bingjiang Campus, which is a public middle school in Jiang'an district of Wuhan, China. This exam is a final English exam for Grade 8 students in Fall semester of 2020-2021 academic year. There are 16 classes with 858 students taking this exam. The data set consists of 858 examinees ($n=858$) and 70 questions ($D=70$), where the questions are from a single exam. Among 70 questions, there are four major types pf questions (Listening Comprehension,  Multiple Choice,  Cloze Test, and Reading Comprehension). We proceed by carrying out an exploratory data analysis. By looking at the estimated accuracy marginalized for each question (column) or each row (examinee), visualized on the left hand side of Figure \ref{fig: expAnalysis}, it is trustworthy that the questions are designed hierarchically in terms of their difficulty, indicated by the estimated accuracy that ranges from 0.247 to 0.981. In addition, the proficiency of examinees are fairly heterogeneous, as the displayed histogram demonstrates a left skewed feature with a long tail. To be more specific, the histogram implies that most examinees can solve more than $70\%$ of the questions, while a small proportion of the examinees, whose estimated accuracy is below 0.4, may have probably failed the test. Such heterogeneity can also be viewed from the boxplots of the Rasch parameters, as shown on the right hand side of Figure \ref{fig: expAnalysis}, where $\xi$ and $\psi$ are defined similarly as those in \eqref{RaschModel}. As the primary goal of analyzing this data set is to explain the \textit{superiority} phenomenon, we move on to the next section to present the results by applying our proposed model.

\subsection{ACBM analysis}\label{sec: ACBM analysis}
To apply our proposed model, we set the number of iterations to be $n_{iter}=400$, $n_{rep}=400$, which are sufficient to thoroughly explore the posterior high density region based on our simulation analyses. The hyperparameters are chosen as $a_0 = 0.01$ and $b_0 = 0.01$ to ensure non-informative prior knowledge while new column and row clusters can still be generated. Given such settings, our model is implemented repeatedly for 100 times with different initial values. The reported column (question) partition is believed to be representative as the median Rand Index between it and the other column partitions is 0.91 with a standard deviation of 0.04 over the 100 Monte Carlo replications. The estimated accuracy parameters using posterior mean under each column cluster and the number of entries corresponding to each accuracy parameter are summarized in Table \ref{tab:3}.
\begin{table}[htp]
    \centering
        \caption{The number of components ($K$) and the estimated component-wise accuracy parameters under each question cluster and the corresponding cluster size ($|c|$).}
    \label{tab:3}
   \begin{tabular}{l|l|l|l}
   \toprule
    \toprule
Cluster  & Size ($|c|$) & $K\leq (|c|+1)/2$ & Estimated accurarcy \\
    \midrule
  1 & 4 & 2 & 0.443, 0.872 \\
2 & 20 & 5 & 0.001, 0.344, 0.645, 0.924, 0.999 \\
 3 & 5 & 2 & 0.218, 0.671 \\
 4 & 17 & 4 & 0.179, 0.419, 0.726, 0.937\\
 5 & 12 & 4 & 0.001, 0.417, 0.752, 0.989\\
 6 & 8 & 3 & 0.164, 0.853, 0.999\\
 \midrule
 7 & 2 & 1 & 0.411 \\
 8 & 1 & 1 & 0.791 \\
 9 & 1 & 1 & 0.247 \\
  \bottomrule
  \bottomrule
    \end{tabular}
\end{table}

Following Table \ref{tab:3}, a question cluster that contains more questions tends to possess more components. The estimation of most component-wise accuracy parameters is precise, since most estimated standard deviation values are one order of magnitude smaller than the corresponding estimated accuracy parameters. We further conjecture that the questions that are assigned to the clusters below the middle line in Table \ref{tab:3} are not effective in distinguishing different types of examinees, suggested by our model. Recall that the maximal number of examinees' mixtures is bounded by $(|c|+1)/2$ to ensure model identifiability. It is hence impossible to identify more than 1 examinees' mixture when a question cluster only has less than 3 questions. In other words, an ideal question cluster should consist of at least 3 questions to be able to detect the heterogeneity among examinees based on the mixture of accuracy model (Binomial mixture model).

\begin{figure}[htp]
\minipage{0.49\textwidth}
  \includegraphics[width=\linewidth]{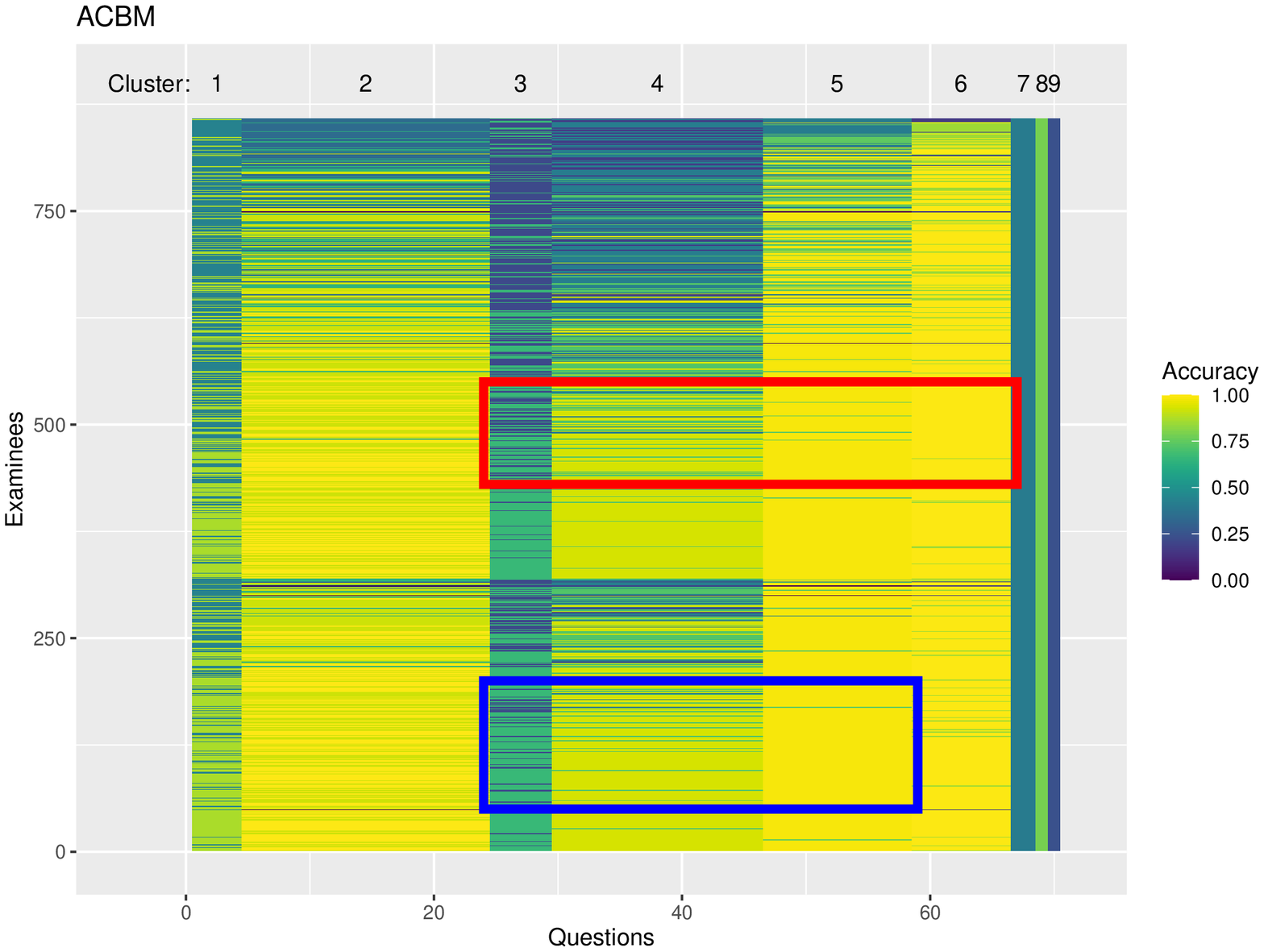}
\endminipage\hfill
\minipage{0.49\textwidth}
  \includegraphics[width=\linewidth]{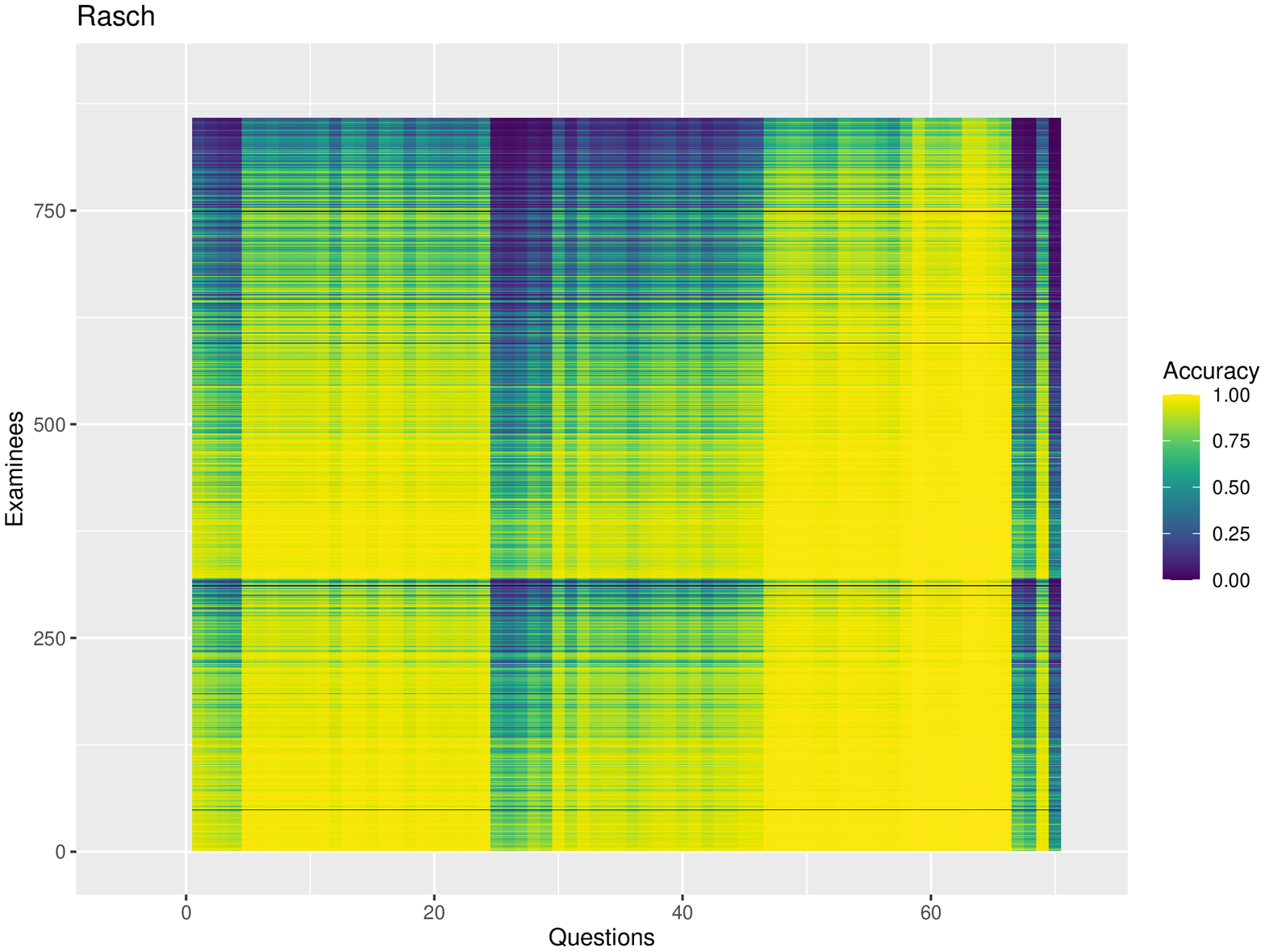}
\endminipage\hfill
\caption {\label{fig: RealPMat} The estimated accuracy parameters aligned in a matrix after permuting the questions based on the estimated column (question) partition. Left: ACBM; Right: the Rasch model.}
\end{figure}

The estimated component-wise accuracy parameters can further be visualized using Figure \ref{fig: RealPMat}, after rearranging the columns (questions) into a consecutive layout according to the estimated column (question) partition given by ACBM, for both ACBM and the Rasch estimations simultaneously.

Intuitively, the gradient plot of ACBM looks like a discretized version of the Rasch gradient plot, which implies that our proposed model can recover the Rasch model's result to some extent. As an advantage over the Rasch model, our proposed model can automatically identify possible question clusters and the mixing structures on the examinees thereof. Note that the Guttman pattern is revealed locally if we look into the accuracy parameters of the corresponding examinees in question cluster 3 and 5. The questions in cluster 5 are listening comprehension and multiple choice questions, which are in general easier compared to the questions assigned to question cluster 3, of which the majority are difficult reading comprehension questions. To provide more insights, we present a contingency table for clusters 3 and 5 in Table \ref{tab:4}. For example, among the examinees who correctly answered to questions from cluster 3 with a higher accuracy (0.671), only 1 of them answer questions from cluster 5 with an accuracy being less than or equal to $.417$. In contrast, 469 ($94.4\%$) of them  answer  correctly to the questions in cluster 5 with an accuracy of .988. This finding agrees with the prior belief that questions in cluster 5 are easier than these in cluster 3, and further indicates that our method can effectively capture the Guttman pattern locally based on specific question clusters.
\begin{table}[htp]
    \centering
        \caption{Contingency table of examinees' count in terms of the accuracy parameters for cluster 3 and 5.}
    \label{tab:4}
   \begin{tabular}{l|l|l|l|l}
   \toprule
    \toprule
\backslashbox[48mm]{C3 (Difficult)}{C5 (Easy)} & Acc = 0.001 &  Acc = 0.417 & Acc = 0.753 & Acc = 0.988 \\
    \midrule
  Acc = 0.218 & 5 & 34 & 73 & 239 \\
  \midrule
  Acc = 0.671 & 0 & 1 & 27 & \textbf{469} \\
  %\midrule
  %\midrule 
 % \backslashbox[48mm]{C3 Acc = 0.671}{Percentage} & 0 & 0.002 & 0.054 & \textbf{0.944}\\
  
  \bottomrule
  \bottomrule
    \end{tabular}
\end{table}

We further discuss the \textit{superiority} phenomenon as indicated by the red rectangle in Figure \ref{fig: RealPMat}. Based on the contingency table \ref{tab:5}, for these examinees who are less proficient in question cluster 3 (accuracy = .218), 54.3$\%$ (25/46) of them did well in question cluster 4 with an 0.937 accuracy. %we find that around 54.35$\%$ of the examinees who are relatively not proficient in question cluster 3 are relatively expert in question cluster 4, suggested by the 0.937 accuracy. 
On the other hand, for those who do well in question cluster 4 (accuracy $\geq$ .726), 37.5$\%$ (42/112) of them did not perform well in question cluster 3 (accuracy = .218).%there are around  the examiness who are relatively poor at (accuracy $\geq$ 0.726) the questions in question cluster 3 performing relatively good in question cluster 4, indicated by the 0.671 Accuracy. 
Such heterogeneity is hardly raised by randomness as we have sufficiently large number of samples in estimating each accuracy parameter. Similar findings can also be discovered in the region formed by the blue rectangle in the same figure. It is attractive to see that our method can capture such \textit{superiority} phenomenon, whereas the Rasch's model is unable to do so by using a single parameter to model the ability of   examinees over all questions.
\begin{table}[htp]
    \centering
        \caption{Contingency table of the count of the examinees in the red rectangle in terms of the accuracy parameters for cluster 3 and 4.}
    \label{tab:5}
   \begin{tabular}{l|l|l|l}
   \toprule
    \toprule
\backslashbox[35mm]{C3}{C4} & Acc = 0.419 &  Acc = 0.726 & Acc = 0.937 \\
    \midrule
  Acc = 0.218 & 4 & 17 & 25 \\
  \midrule
  Acc = 0.671 & 3 & 14 & 56  \\
  \bottomrule
  \bottomrule
    \end{tabular}
\end{table}

\section{Discussion}\label{sec: Discussion}
In this article, we propose a novel IRT model using averaged mixture of binomial distributions with constraints, of which the novelty basically comes from the modelling of the \textit{superiority} phenomenon and the justification of the identifiability issue. Our model is manifested to be effective in both theoretical and practical aspects. Namely, the identifiability conclusion and posterior contraction results indicate that the latent accuracy parameters of our interest can be estimated at a $\sqrt{n}$ (up to a $\log$ term) rate asymptotically. In addition, the posterior samples of these parameters can be obtained using a tractable sampling algorithm, which satisfyingly approaches the stationary distributions according to the simulation results. By exploring the clustering effect at question level, depending on which the heterogeneity among examinees is further introduced, the real data analysis results given by our model are more flexible and interpretable compared with the Rasch model and many existing methods.

One possible generalization of our model is to consider the product of Bernoulli densities in place of the Binomial density, such that the accuracy parameters of the questions assigned to a question cluster can be arranged in an ascending order after a permutation. In other words, without loss of generality, suppose there exists a permutation $\sigma(\cdot)$ given a question cluster indexed by $1,2,\ldots,D'$, we define the product of Bernoulli densities as
\begin{equation}\label{ascending}
    \begin{split}
        & X_{i,j}\stackrel{ind}{\sim}\text{Bernoulli}(p_{i,j}), ~ p_{i,\sigma(1)}\leq p_{i,\sigma(2)}\leq \ldots \leq p_{i,\sigma(D')},~\text{for}~j=1,\ldots,D',
    \end{split}
\end{equation}
where $\sigma(\cdot)$ is shared within the question cluster. We may call it the ordered Bernoulli densities. We expect such a gradient of the accuracy parameters can better explain the Guttman pattern than the kernel function currently used. The only concern of this structure is the identifiability of using this kernel density, which requires further investigation. One can directly apply our theoretical results if the product of Bernoulli densities is shown to be first-order identifiable under certain conditions. Another possible way of improving is to consider a more advanced sampling algorithm than ours, which is a typical application of the Algorithm 1 proposed by \citet{neal2000markov}. One may use split-merge sampling \citep{jain2004split} or slice sampling \citep{neal2003slice} to boost the procedure of approaching the stationary distribution. Future simulation studies would also be attractive by comprehensively examining the difference between our model and the alternatives that accommodate heterogeneity by introducing mixing structure at both the item and subject levels.

\bibliographystyle{chicago}
\bibliography{ref.bib}

\end{document}